# Blockchain as an IoT intermediary

Matija Šipek [1-2], Martin Žagar [2] and Branko Mihaljevic [2]

[1] CS Computer Systems, Zagreb
[2] RIT Croatia, Zagreb

**Abstract.** Blockchain technology provides a private, secure, transparent decentralized exchange of data. Also, blockchain is not limited to a particular area, but it has a wide range of applications and can be integrated into a variety of Internet interactive systems. For example, the Internet of Things (IoT), supply chain tracking, Electronic Health Records (EHR), digital forensics, identity management, trustless payments, and other key business elements will all benefit from its implementation. Next layer solutions such as Ethereum 2.0, Polkadot, Cardano, and other Web 3.0 technologies provide developers versatility. Moreover, these platforms utilize smart contracts which are similar to standard, traditionalized software during development but offer key utilities to end-users such as online wallets, secure data with transparent rules.

Blockchain is receiving a lot of attention in educational technology (EduTech) as it aims to achieve a more transparent and multipurpose educational system. In addition to smart contract technology which defines how data should be registered, gathered and processed, blockchain can be used as an IoT intermediary for mobile usage.

Therefore, we implemented an educational learning platform powered by blockchain technology to examine feasibility in industry and academic environment. In essence, this is a web application which is adapted to mobile platform and connected to blockchain for crucial data exchanges. In this paper we want to emphasize the potential of blockchain technology in multiple sectors as well as the need to really understand the underlying principles which are allowing disruptability of traditional centralized software solutions.

## 1 Introduction

In this paper, we are presenting a decentralized system powered by blockchain technology also used as an IoT system. While there are research articles that use blockchain as an IoT intermediary [1] most of them do not go into enough detail regarding needed technical specification as there is no one-size-fits-all. Every blockchain network has the *blockchain trilemma*: decentralization, security, and scalability. These factors are interconnected, meaning if you try to improve one facet, another one will decrease as a trade-off.

Due to a lack of transaction speed some IoT blockchain systems [2] tend to eliminate particular parts, for example removing consensus mechanisms which just makes the need of a blockchain completely redundant and without the needed level of security.

On the other hand, in order to achieve a similar level to legacy systems of transaction processing, blockchain networks must be capable of handling an enlarging number of users, transactions and critical data. For instance, a common comparison is Bitcoin which can handle around 4.6 transactions per second while Visa handles around 1700. The problem is obvious, if blockchain networks were to compete with legacy systems they need to at least match the current numbers.

## 1.1 Centralized system

Most of today's internet systems and applications are centralized, meaning that a central authority is in control of data and functions an individual is using. An example would be an email service; the provider knows when you send the message and to whom, the data is allegedly privately stored, still, the email service has a copy of your data to which it is not impossible to access.

It can be claimed [3] that a centralized system offers better security, as there is less surface of attack compared to decentralized systems; however, there is always a possibility of a *coup d'état* whereas central entity goes rogue and maliciously uses data for its own benefit, also not an uncommon event.

Traditional database systems can have a reasonable amount of privacy and security with additional measures like firewalls, but the centralized nature and lack of user control raise concerns about the true level of privacy.

Still, it can be said that centralized systems are more efficient since there is a clear command structure; it is simpler to plan since network and infrastructure requirements can be predetermined, and quick changes and decisions can be made.

## 1.2 Decentralized systems

The widespread adoption of blockchain and distributed ledger technology has resulted in a workable solution that presents a consensus of synchronized digital assets that are not owned by a single entity and are instead stored on a public digital ledger.

The shift from a centralized to a decentralized approach to application deployment opens up new possibilities for several industrial disciplines where access management, integrity, immutability, and non-discriminatory governance of data is imperative.

The key safety mechanism is a consensus protocol and the most common type of this verification is Proof-of-Work (PoW), used by the majority of modern cryptocurrencies. The process is known as mining, and the construction of a new block is completed by solving a complex mathematical puzzle that yields a 256-bit number hash, which is a unique identifier of data within the block.

Furthermore, the distribution of computational power and storage is an important security feature of a blockchain. As a result, because each node preserves a complete replicated transaction history, so if an attacker tries to profit through a double-spend attack, a 51 % attack, or any other malevolent incentive, he will need an enormous amount of resources in relation to the respected consensus algorithm.

Even so, the high level of security PoW provides is diminished by the high resource consumption and slow transaction speed, making it unfeasible for large-scale adoption.

## 1.3   Layer-1

Layer 1 blockchain is a set of solutions that are incorporated in the blockchain's core protocol to improve the blockchain's functionality and scalability. Consensus protocol modifications and sharding are the two most popular Layer-1 scalability solutions.

The new consensus protocol Proof-of-Stake (PoS) tries to remove costly mining operations with the introduction of *Staking*. To be able to participate in the verification and creation of blocks on a blockchain, you must stake tokens. In a Proof-of-Stake blockchain, an individual or group is picked at random to verify transactions. The algorithm weighs the number of tokens they have staked on the network as collateral. Those chosen to confirm a block are routinely rewarded with the transaction fees associated with that block. The stake serves as a deterrent to malicious behavior.

Sharding is a way of splitting the database horizontally into smaller distinct datasets called "shards". This way, not all nodes need to maintain a copy of the entire network, thus allowing parallel processing so sequential work can be done on numerous transactions.

## 1.4   Layer-2

Layer-2 solutions refer to the collective pool of technologies designed to run on top of an underlying blockchain network in order to enhance its efficiency. The base layer blockchain becomes less congested and ultimately more scalable by abstracting the majority of data processing to auxiliary networks. A great example of this is *Rollups,* the heavy-computational transaction execution is performed outside the main chain, but data is posted on Layer-1, meaning its secured by Layer-1 and also achieves much better performance. There are several similar concepts in Layer-2 sphere: Nested Blockchains, State Channels, and Sidechains, and not to go into too much detail, these are all trying to fix the limitations of layer-1 technologies [4].

## 2   Internet learning platform

Smart contracts are a key component of programmable distributed ledgers like Ethereum, Cardano, and Polkadot. Smart contracts are programs that run on the global blockchain; the code, as well as all of the data controlled within the transactions, is public, resulting in a system that is trustworthy and cannot be cheated if properly designed. In our system implementation, we took into consideration the lack of scalability and designed our system to handle transactions in a similar way to Layer-2 Rollups; namely, separation of concerns so that only light-weight end result data is put on-chain.

Main communication with the user is done on a simple website and all the heavy computation is done on a separate .NET API platform as can be seen from Fig.1.

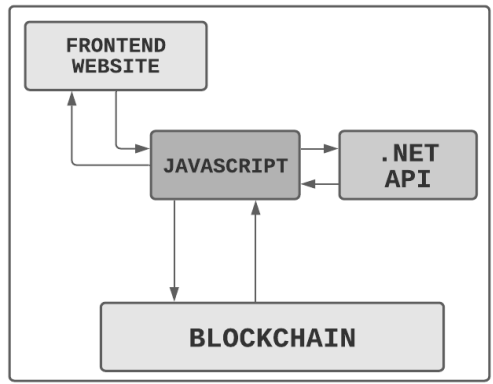

**Fig. 1.** High-level system overview

### 2.1 Ethereum network-based solution

In our initial system design, we selected Ethereum Virtual Machine (EVM)-based system replica to emulate the real network. Our architecture connects three main sectors with clear duties that are linked by an intermediary level that manipulates and shares data. In order to replicate EVM, we had to employ multiple technologies: Truffle suite, Ganache, Web3.js, MetaMask, Solidity and other technologies.

*Truffle suite* is a development environment that includes smart contract compilation, linking, and deployment to different networks, as well as an interactive debugger, numerous libraries, and automated testing, as well as scriptable deployment and migration frameworks.

*Ganache* is a local in-memory blockchain environment that mimics the behavior of distributed ledgers in the reality. Solidity-related files, artifacts, and migrations are the three fundamental pieces of the execution model that split smart contract functionalities. The system's starting point is composed of solidity contracts, solidity-related libraries, and dependencies.

*Web3.js* is a suite of libraries that allow communication with Ethereum nodes, both local and remote, using various network protocols like HTTP, IPC, and WebSocket.

*MetaMask* is a browser plugin that acts as a crypto wallet, allowing websites to request Ethereum accounts and therefore run Ethereum decentralized applications (dApps).

*Solidity* is an object-oriented high-level language for smart contract development. A smart contract is a collection of code, functions, variables, and state that is recorded and exists on the blockchain at a specific address.

**2.2 WIP Cardano network-based solution**

The current state of smart contract implementation in the Cardano network can be defined as development in progress [5]. The team behind Cardano is implementing the Plutus Platform for running smart contracts both on-chain and off-chain code, and the update is named the 'Alonzo' hard fork. Recently, an alpha test net, Alonzo Blue, has been launched to a testing community with exercises in order to assess solutions with incorporate next areas: Hard Fork, Addresses and Transactions, Managing Native Tokens, Interaction with the Wallet etc.

While all of this sounds great, the key factor of compiling and submitting Plutus scripts is not implemented yet, meaning our tests on the Cardano network will have been marked for future work.

**2.3 WIP Polkadot network-based solution**

Polkadot is a next-generation blockchain technology that can connect numerous specialized chains in a universal network, with a focus on building Web 3.0 infrastructure. When comparing Polkadot to other blockchains, several new concepts are introduced, most importantly *Parachains* and *Parathreads* and it changes the way smart contracts are interrelated with these [6]. The main Polkadot Relay Chain does not natively support smart contracts, however, the concept of Parachains will allow for more precise and fast execution of extensive logic than a smart contract platform could provide. For example, Custom fee and Monetary Policy structure, Decentralized Autonomous Organization (DAO) Governance, Shared Randomness etc. Arguably, with a more complex runtime logic, an abstraction layer is added on top of it as developers will be able to declare a whole environment of their chain, also allowing smart contracts to be written on top of them [7]. Both, Polkadot and Cardano present potential future benefits based on the concept of om- and off-chains codes in universal networks, implementation is still not possible due to the lack of network details. That's why we proceeded with the implementation described in Section 2.1.

## 3 Results

Our educational learning platform comprises several interconnected technologies which together create a useable, precise, and responsive system with agility for further operations. One of the main goals was to focus on privacy guarantees for both students and professors such as decentralization and differential privacy.

We have built our educational learning platform based on smart contracts to examine all potential benefits and flaws in the technology. In order to be able to use smart contracts, we compiled and deployed them using contract creation transactions. In this process, each contract is associated with an address derived from the creating transaction, more specifically from the invoking account and the nonce, which is a timestamp. After deployment, contracts are ready to be called and run by externally owned accounts (EOAs), in our case EOA's calls will be users, that will call smart contracts after each simulation iteration circle finishes. Our system uses solidity v0.5.0 with which we are deploying contracts to the Ethereum blockchain. The key communication line for blockchain reporting is the web3.js connected to a Solidity object.

Our learning platform simulates a simple business environment of IoT on Blockchain where students (representing nodes in IoT) are supposed to simulate digital marketing efforts for a hypothetical smartphone brand with a limited range of three devices with different technical specifications and a targeted market. The simulation consists of several rounds in each of which students are planning the budgets per digital platform and decisions about other parameters (e.g., keywords strategy). After each round, activity reports (i.e., digital platform insights/analytics) as a result of execution of smart contracts are generated, providing students with valuable feedback on their actions. During the start of the simulation, pre-determined data entered by the admin is loaded on the blockchain. This data is the starting benchmark of globally defined values which define successfulness of each user's overall situation e.g., Number of likes, Post engagement, Page views, etc. and they are equally set for all users.

Since the platform is based on the concept of a smart contract, all the activity reports are immutable, showing data integrity and enabling everyone in the simulation is dealing with the same datasets. The simulation consists of 16 iterations where the initial iteration is for instructional setup and creating the environment. Reporting at the end of each round is based on smart contracting and for executing the contract, it is required to pay a gas fee. Our basic solution is built on the previously mentioned EVM-based environment with Truffle suite, Ganache, Web3.js, and MetaMask, and smart contracts developed in Solidity. In order to compare our proposed solution on Ethereum, with possible Cardano and Polkadot solutions, we adjusted the predicted amount of activities per iterations of each possible solution.

As a major benefit of our solution, we have proved that our solution is applicable for use in any kind of smart contract. The time to the finality of executing the smart contracts on our platform is presented in Fig. 2. It could be seen that the average time to finality is denoted in a couple of seconds which is acceptable time for proving and executing business contracts, comparing to the real world where such operations could take even a few days.

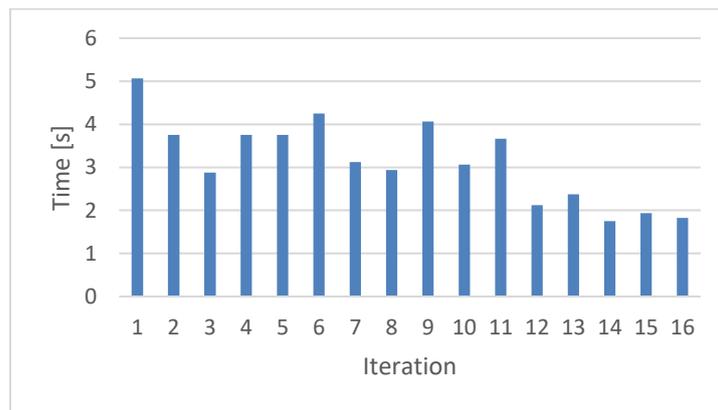

**Fig. 2.** Time to finality per simulation iteration on our platform

As an additional outcome, as presented in Fig. 3 on average transaction costs, while our solution based on Ethereum is constantly showing costs around 0.7 gas (which is currently 15.80 Gwei), predicted gas price based on [8] and [6], for reporting based on Polkadot and Cardano technologies will be some 3 times lower while enabling even broader and cheaper usage of our platform.

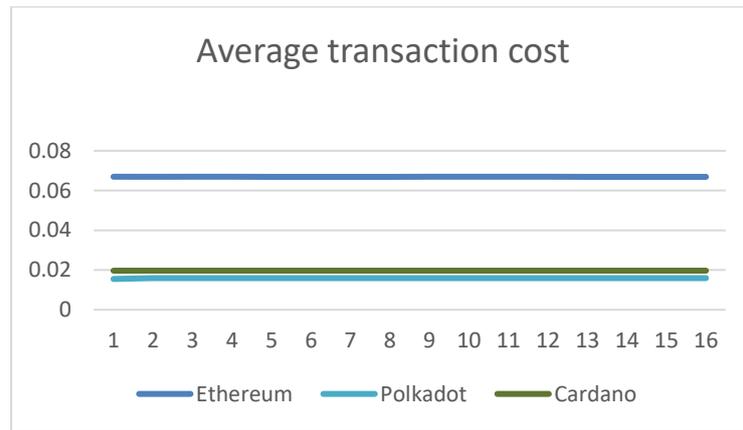

**Fig. 3.** Comparison of average transaction costs in gas

## 4     Conclusion

According to revised Bloom's taxonomy, the creation of this new and original educational learning platform, for the educational process and proving Blockchain technologies as IoT intermediaries (in our case, students represented nodes in IoT) is able to cover all six cognitive processes (which is in our case valid for both Business and IT students). Business students get an insight into strategic knowledge and conceptual knowledge by using the platform, and knowledge of Internet marketing specific techniques and methods, knowledge about the terminology, together with knowledge about the criteria for determining when to use appropriate procedures when designing business logic in the background of this application. IT students acquire knowledge of distributed application development skills and Blockchain and smart contract technologies, knowledge about the classification when acquiring the initial input data from Business students, and knowledge about the terminology in building the distributed applications from their instructor.

We also proved that our approach based on EVM is applicable to different business domains that is based on the immutability and data integrity provided by the smart contracts. We anticipate gas fees as the only downside, although the relevant numbers are counted in cents, which would not be a problem in real business application for replacing and approving the business contracts with the smart contracts.

Our result datasets show that this new approach of gamification in learning platforms is inclusive towards different perspectives of knowledge sharing and knowledge gaining, and in the near future when other frameworks like Cardano and Polkadot will be fully available, with additional options and lower gas prices and off-chain code management in universal networks, our proposal could be defined as a benchmark for educational learning platform for business and IT curricula, and also a showcase for proving the applicability of smart contracts in business applications.